# Mine Blood Donors Information through Improved K-Means Clustering

Bondu Venkateswarlu[1] and Prof G.S.V.Prasad Raju[2]

[1]Department of Computer Science and Systems Engineering, Andhra University, Visakhapatnam-530003,India.

iambondu@yahoo.com

[2]School of Distance Education, Andhra university, Visakhapatnam-530003,India.

gsvpraju2011@yahoo.com

## Abstract

*The number of accidents and health diseases which are increasing at an alarming rate are resulting in a huge increase in the demand for blood. There is a necessity for the organized analysis of the blood donor database or blood banks repositories. Clustering analysis is one of the data mining applications and K-means clustering algorithm is the fundamental algorithm for modern clustering techniques. K-means clustering algorithm is traditional approach and iterative algorithm. At every iteration, it attempts to find the distance from the centroid of each cluster to each and every data point. This paper gives the improvement to the original k-means algorithm by improving the initial centroids with distribution of data. Results and discussions show that improved K-means algorithm produces accurate clusters in less computation time to find the donors information.*

## I Introduction

Now a day, due to large number of accidents and health problems, it is dire need to find the blood donor information instantly. Since the blood bank repositories are rapidly growing, it is becoming increasingly difficult to extract the information required by using the conventional database techniques. So, there is a need to come up with a solution using clustering for analysis and obtain the information of the blood donors [1].

Clustering analysis [2] is one of the major data analysis methods, which help to identify natural grouping of data objects from the dataset. Clustering is unsupervised classification and is a process of partitioning a set of data objects into a set of meaning full sub classes. This can be done by applying the various similarity and distance measures to the algorithm.

For all geometrical problems, the standard distance metric used is the Euclidean distance, which is just the direct distance between any two points and can be easily measured with the help of just a ruler, be it in two-dimensional space or three-dimensional space. It is a default distance measure [3] for the k-means clustering algorithm.

K-means Clustering Algorithm is a Partitioning algorithm; it is effective in producing for many practical applications [4, 5, 6, and 7] like Pattern recognition, Document classification, Image processing, Economic science. However, while calculating the initial cluster centroids, the K-means algorithm is vulnerable to errors. Because of this the initial cluster centroids are produced arbitrarily, hence disabling the algorithm from providing the expected results. The accuracy of the initial centroid determines the efficiency of the original k-means algorithm. This paper attempts to improve the accuracy of the k-means algorithm by increasing the accuracy in the calculation of the initial centroids.

## II *K-Means Clustering Algorithm*

The K-means algorithm assigns each point to the cluster whose centroid (Centre) is in the nearest proximity (centroid is the average of all the points in the cluster). The centroid's coordinates are determined by calculating the arithmetic mean of all the points in the cluster separately for each dimension. The biggest asset of this algorithm is its speed, which allow for it to be run on large databases. Also the simplicity of this algorithm makes it all the more easy to use. But this algorithm has its own share of shortcomings associated with it. The inability of this algorithm to yield the same results with each run is a major disadvantage and this occurs as the derived clusters are not independent of the initial random assignments. Also, this algorithm considerably reduces the intra-cluster variance, but does not take measures to ensure the global minimum of the variance of the result.

There are two separate phases in the k-means algorithm- wherein the first phase involves the identification of k centroids, given that we know the number of clusters (K) beforehand thereby arriving at one centroid per cluster. Initially K points which are likely to be in different clusters have to be selected, which are then made the centroids of their respective clusters. The Euclidean distance is then calculated for each data point from each of the cluster centroid. Compare the values; find the closest centroid for the data point. Then bind the centroid and the data point which results in the completion of the first phase and then an early grouping is done.

At this stage, the new centroids have to be determined by calculating the mean value for each cluster. After we get k new centroids, then a new binding is to be created between the same data points previously used and the nearest new centroid, thus generating a loop. Because of this loop, the k centroids are prone to a change in their positions in a step by step manner. This process is repeated until convergence criteria is meat means the centroids of clusters are do not move anymore. The pseudo Code for the k-means Clustering algorithm.

**Algorithm** : The k-means clustering algorithm

Input:

D = {d1, d2,......, dn} //set of *n* data items.

*k* // Number of desired clusters

Output:

A set of *k* clusters.

Steps:

1. Arbitrarily choose *k* data-items from D as initial centroids;

**2. Repeat**

2.1 Assign each data item *di* to the cluster which has the closest centroid;

2.2 Calculate the new mean of each cluster;

**Until** convergence criterion is met.

## III *Literature Survey*

In a web based information system for blood donation available on www. Pavel Berkhin[11] performed extensive research in the field of data mining experiments and organized analysis of the blood bank repositories which is helpful to the health professionals for a better management of the blood bank facility.

Arun K.Pujari et al. [12] have worked to improve the performance of blood donation information analysis. In this paper, improved k-means clustering is adopted that improves the performance for determining the blood donors information based on the required blood group and location where needed.

Several attempts are made by the researchers to improve efficiency of the k-means clustering [8,9,10]. The variants of the k-means clustering algorithm, are K-modes and K-medoids. These algorithms replace the means with the modes and medoids. The K-modes algorithm handles the categorical data.

These also give better performance based on the way we choose the initial modes and methods. For clustering the data, the k-means and the k-modes are integrated by the k-prototypes algorithm. Both the numeric and categorical attributes are taken into account to define the dissimilarity measure.

Fahim A.M. et. al.[8] and Yuan F et.al. [10] worked for improving the performance of the k-means by finding the initial centroids. The centroids obtained this by these methods provide consistent and accurate clusters for the given datasets.

Abdul Nazeer K A et.al [13] brought forward an efficient method for assigning the data-points to the clusters. The original k-means algorithm it computes the distances between the data points from all the centroids in every iteration, thus making this algorithm computationally very expensive. Another approach given by Fahim makes use of two distance functions for this purpose- one similar to the k-means algorithm and another one based on a heuristics to reduce the number of distance calculations. However as was the case in the original k-means algorithm, this algorithm too determines the initial centroids randomly, thus compromising the accuracy of the resultant clusters.

Koteswararao[2] proposes improved k-means algorithm using a $O(n \log n)$ heuristic method for finding the initial centroids. In this method, the initial centroids are generated in accordance with the distribution of the data instead of generating them randomly.

# IV *Proposed Methodology*

Find the blood donors information and satisfy the need by these steps:

Collect data from the blood bank.

Apply the improved K-means Algorithm to classify the number of blood donors through the blood group and location where it needed.

Extract the Mail ids from the resultant cluster which satisfies the criteria of blood group and location. Forward the message to the donors.

In this methodology, the improved clustering algorithm, deals with the multi dimensional data values. Each data point *di* may contain multiple attributes such as *di1, di2,…..dim*, where *m* is the number of attributes or columns in each data value. In such cases we first determine the column with maximum range [7], where range is the difference between the maximum and the minimum element in the column.

Then determine the initial centroids [10] as range of the Column is divided by the number of clusters and sum with the minimum value of the column. i.e., assume we have the two dimensional dataset, then the initial centroid are

$Cx = MAXx - MINx K+1 + clusterid * MINx$  $Cy = MAXy - MINy K+1 + clusterid * MINy$

After determining the initial centroids, the data points are distributed to the each cluster as specified by the Abdul Nazeer [13]. That is the data points are divided into k-equal partitions.
For each iteration, calculate the Euclidean distance between the data point and each centroid.

For all geometrical problems, the standard distance metric used is the Euclidean distance, which is just the direct distance between any two points and can be easily measured with the help of just a ruler, be it in two-dimensional space or three-dimensional space. It is a default distance measure [3] for the k-means clustering algorithm. Clustering problems and clustering text widely use of Euclidean distance, which is a true metric as it satisfies all the above stated four conditions. Also the Euclidean metric is the default distance metric which is used in the k-means algorithm. The distance measured by the Euclidean metric is also known by 'as-the-crow-flies' distance. This distance from a point $X$ ($X_1, X_2,$ etc.) to a point $Y$ ($Y_1, Y_2,$ etc.) is:

The square root of the sum of the squares of the differences between the corresponding values of the two data points is necessary for finding the Euclidean distance.

Find new centroid by calculate the mean value of distance of the all data points of that cluster. Each data point $d_i$ is assigned to the cluster having the closest centroid. The distance between the data points and the centroid is measured by using the Euclidean distance. Improved K-means Clustering Algorithm is outlined as below

**Algorithm 2**: K-means Clustering Algorithm with improved initial centroids.
**Input:**
D = {d1, d2,......,dn} // set of *n* data items.
*K* // Number of desired clusters.
**Output:**
A set of *k* clusters.
**Steps:**
Calculate the initial centroids as the formula give above and set the cluster with that centroid.
Repeat
        2.1 Initially assign the each data point to the cluster
        2.2 Update the centroid value by calculate the mean of that cluster
**Until** all data points are assigned to any one of the clusters.

3. **Repeat**
3.1 Assign each data item *di* to the cluster which has the closest centroid;
3.2 Calculate new mean of each cluster;
**Until** convergence criterion is met.

This algorithm gives the better performance and accurate results within less time compared to the literature [10,12].

## V Results and Discussions

For testing accuracy and efficiency of original K-means and improved K-means clustering algorithm, various datasets are used. Data sets are varying in size. The same data sets are given to the original K-means and improved K-means algorithm.

Both algorithms require the inputs as number of clusters, and the required blood group and location where it needed. The original k-means algorithm takes the initial centroids as the arbitrarily choosing random values; the improved k-means algorithm takes the initial centroid in more meaning full way by calculated using formulas.

The efficiency of this algorithm can be analyzed by the accuracy of the clusters obtained from the experiments performed on the blood bank dataset which satisfy the given criterion (here blood group). The time taken and the percentage accuracy for each experiment are computed and tabulated. Performances of the algorithms for the different data sets are tabulated in Table I. The results produced from different algorithms are compared in figures 1 to 2.

Figure1 describes the accuracy of the algorithms, y-axis shows the accuracy. Figure 2 describes the time taken by the algorithm, y-axis shows the time taken by the algorithm.

The performance of the improvised algorithm as measured against the original one is tabulated in the Table 1. It shows the results for different data sets.

Table 1:PERFORMANCE COMPARISION OF THE ALGORITHMS FOR DIFFERENT DATASETS BY ACCURACY

| Datasets | Original K-means | Improved K-menas |
|---|---|---|
| >10000Records | 79.2 | 83.4 |
| >5000Records | 86.5 | 90.2 |
| >1000Records | 96 | 96.5 |

Figure 1

Table 2 shows the time which the improvised K-Means takes to obtain the data from different datasets as measured against the Original K-Means.

Table 2:PERFORMANCE COMPARISION OF THE ALGORITHMS FOR DIFFERENT DATASETS BY TIME

| Datasets | Original K-means | Improved K-menas |
|---|---|---|
| >10000Records | 40 | 28 |
| >5000Records | 30 | 17 |
| >1000Records | 7 | 3 |

Figure 2

## VI *Conclusion*

This K-means algorithm has many real time applications, but its performance cannot be guaranteed as it takes the initial centroids randomly. Also the computational complexity of the original k-means algorithm is alarmingly high considering the need to reassign the data points many number of times, as they are reassigned every time the loop runs. So that we are taking the initial centroids in a meaning full way and assign data points to the clusters in a distributed manner. These results in better accuracy compared to the classic K-means Algorithm. Results and

Discussions show that the improved initial centroids with k-means clustering algorithm give the accurate and efficient results. So that we fulfill the need of the many people by forward the mails to the blood donors.

A limitation of the improved k-means algorithm is , it still takes the input as number of clusters and take the location in a static way. For future work for this paper is place the GPRS positioning system and take the number of clusters by the size of the dataset.

## *Authors*

[1]Venkateswarlu Bondu received the Master's Degree in

Computer Science and Systems Engineering from Andhra University

College of Engineering, pursuing Ph.D in Department of Computer Science and

Systems Engineering Andhra University,Visakhapatnam.

His research area is Data Mining and Data Warehousing.

[2]Prof G.S.V.Prasad Raju Professor in Computer Science ,in School of Distance Education, Andhra University,Visakhapatnam. His area of specialization is E-Commerce, Internet Technologies.